\documentclass[titlepage,twocolumn,superscriptaddress,groupedaddress,nofootinbib]{revtex4}

\usepackage[pdftex]{graphicx}
\usepackage{graphicx}
\usepackage{dcolumn}%
\usepackage{amsmath} 
\usepackage[caption=false]{subfig}
\usepackage{color}
\usepackage{hyperref}

\begin{document}
\title{A channel for very high density matter-antimatter pair-jet production by intense laser-pulses}

\author{D. Del Sorbo\footnote{Corresponding author}}
\affiliation{York Plasma Institute, Department of Physics, University of York, York YO10 5DD, United Kingdom}
\affiliation{High Energy Density Science Division, SLAC National Accelerator Laboratory, Menlo Park, CA 94025, USA}
\author{L. Antonelli}
\affiliation{York Plasma Institute, Department of Physics, University of York, York YO10 5DD, United Kingdom}
\author{P. J. Davies}
\affiliation{The School of Physics and Astronomy, University of Manchester, Manchester M15 9PL, United Kingdom}
\author{L. N. K. D\"ohl}
\affiliation{York Plasma Institute, Department of Physics, University of York, York YO10 5DD, United Kingdom}
\author{C. D. Murphy}
\affiliation{York Plasma Institute, Department of Physics, University of York, York YO10 5DD, United Kingdom}
\author{N. Woolsey}
\affiliation{York Plasma Institute, Department of Physics, University of York, York YO10 5DD, United Kingdom}
\author{F. Fiuza}
\affiliation{High Energy Density Science Division, SLAC National Accelerator Laboratory, Menlo Park, CA 94025, USA}
\author{H. Chen}
\affiliation{Lawrence Livermore National Laboratory, Livermore, California 94550, USA}
\author{C. P. Ridgers}
\affiliation{York Plasma Institute, Department of Physics, University of York, York YO10 5DD, United Kingdom}


    \maketitle
   
\textbf{The mechanism of laser-driven relativistic pair-jet production qualitatively changes as laser intensity exceeds $I\gtrsim5\times10^{22}$ W/cm$^{2}$ because of the appearance of laser-induced strong-field QED processes. Here, we show that by exceeding this intensity additional physics operates and opens a new and efficient channel to convert laser photons into dense pair-jets -- the combination of nonlinear Compton scattering and the Bethe-Heitler process. This  channel generates relativistic electron-positron jets more than three orders of magnitude denser than has so far been possible. We find that the process is so efficient that it leads to the prolific production of heavier pairs as well. The jets produced by this new channel will enable the study of collective processes in relativistic electron-positron plasmas.}

Relativistic electron-positron plasmas can be found in extreme astrophysical environments, such as pulsar magnetospheres \cite{cerutti2017electrodynamics} and are central to explain energetic phenomena related to $\gamma$-ray bursts  \cite{mirabel1999sources}. 
Collective effects in these pair plasmas and jets are important. For example, in pulsars, pair discharges generate pair plasmas dense enough  to screen the accelerating electric field, inducing the formation of a `force-free' magnetosphere \cite{cerutti2017electrodynamics}. Producing relativistic electron-positron plasmas of sufficient density and extent that those collective effects may occur in the laboratory will soon be possible.

A diverse range of experimental investigations have been undertaken to study collective effects in non-relativistic electron-positron plasmas.
 For example, two-stream instabilities have been observed in the interaction of an electron beam with a pure positron plasma captured in a Penning trap \cite{greaves1995electron}. In this case the study of collective modes is limited as the annihilation time of the positrons is relatively short and the density of the positron plasma is low ($\sim 10^{6}$ cm$^{-3}$). To circumvent this, partially analogous plasmas containing only ion species of equal mass have been created in fullerenes \cite{oohara2005electrostatic}.  
Positron trapping has been observed in a levitating dipole magnetic field \cite{saitoh2015efficient} and there are plans to magnetically trap an electron-positron plasma using a stellarator magnetic confinement device \cite{pedersen2012plans}.

With the development of high-intensity laser pulses ($I\gtrsim 10^{19}$ W/cm$^{2}$), large numbers of relativistic positrons can be created in a small volume in a time short compared to their lifetime, potentially enabling the study of astrophysically relevant relativistic pair plasmas \cite{chen2015scaling}. Relativistic electron-positron pair-jets have been generated experimentally using ultra-intense lasers as follows: the laser accelerates electrons to $\gtrsim$ MeV energies; these energetic electrons then radiate $\gtrsim$ MeV $\gamma$-ray  photons by bremsstrahlung as they propagate through a heavy target  (i.e~characterized by a large number of nucleons); $\gamma$-ray  photons can then decay to pairs by Bethe-Heitler process \cite{heitler1954aa}, in the strong fields around the target nuclei.  The first step, electron acceleration by the laser, can occur in a separate gas-jet target by laser wakefield acceleration, or acceleration by the laser fields in the plasma created as a laser strikes a solid target directly.  Both techniques have been applied successfully to generate large numbers of electron-positron pairs \cite{sarri2015generation, chen2009relativistic, chen2010relativistic}. Using the latter technique it has been possible to generate relativistic electron-positron jets of density as high as $10^{16}$ cm$^{-3}$.  At present, experiments operate close to where the relativistic skin depth is of the order of the jet size. 
Therefore, 
this prevents the laboratory investigation of many collective phenomena

\begin{figure}
        \centering
        \includegraphics[width=1.\columnwidth]{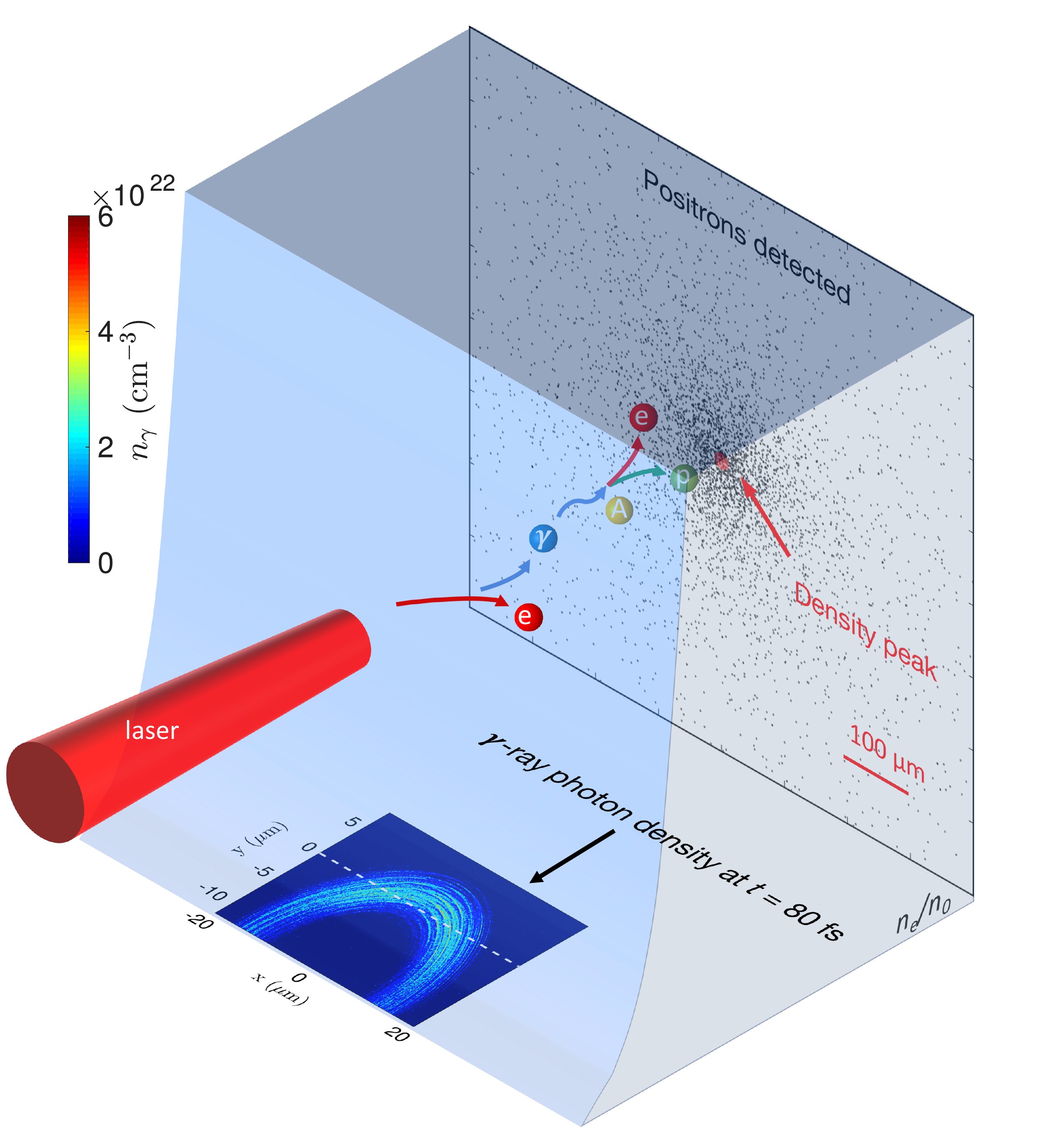}

      \caption{Schematic representation of ultra-intense laser illumination of a gold target. Part of the laser energy is converted into  $\gamma$-ray photons in the preplasma created by the laser prepulse. This conversion is mainly due to nonlinear Compton scattering of electrons in the laser focal spot. $\gamma$-ray photons propagate into the unperturbed gold. Interacting with heavy ions by Bethe-Heitler process, $\gamma$-ray photons create electron-positron (and muon-antimuon) pairs that form a jet at the rear side of the target. The figure also shows the pre-ablated target density $n_{e}$, when the main pulse illuminates the target at 
      $t=0$, the $\gamma$-ray photon density $n_{\gamma}$  induced by a laser with $I=10^{23}$ W/cm$^{2}$ at $t=80$ fs, and the positron detection at the rear side of a $50 \;\mu$m thick target.
\label{disegno}
 }
\end{figure}

 In order to study collective effects a higher density relativistic electron-positron pair-jet is necessary. 
    Near future multi-petawatt laser systems achieving peak intensities of $\sim10^{23}$ W/cm$^{2}$ and above, such as the Extreme Light Infrastructure, should enable us to study these collective phenomena \cite{negoita2016laser,turcu2016high}.  
  The production of pair-jets in heavy targets in this environment remains unexplored and is potentially qualitatively different due to the importance of strong-field QED processes -- such as nonlinear Compton scattering \cite{di2010quantum} 
 in a laser produced plasma
 \cite{del2018efficient,ridgers2012dense,bell2008possibility}.
 
Here we show how the process of generating dense electron-positron jets in laser-solid interactions is expected to change qualitatively in the multi-PW regime.  We prove that in this regime pair-jet production occurs via a new laser conversion channel, characterized by a short mean free path that minimizes the jet divergence.
These jets are produced by a two-step processes whereby a bright flash of $\gamma$-ray photons is produced by non-linear Compton scattering in the laser-plasma interaction at the front of the target.  These photons then decay to pairs via the Bethe-Heitler process in the bulk of the heavy target.
 The generated relativistic pair-jets are three orders of magnitude higher density than currently achievable and
 are sufficiently high such that the relativistic skin depth is several times smaller than the jet size. This difference is significant as it allows for the first time clear emergence of collective phenomena in a laboratory source. 

 \section*{Results}

\textbf{Simulation of dense pair-jet production.} The numerical simulation of the interaction of an ultra-intense laser ($I>10^{22}$ W/cm$^{2}$) with a heavy target, including all the relevant processes, has been achieved by coupling the particle-in-cell (PIC) modeling of the laser-plasma interaction at the front of the target to the Monte Carlo modeling of the subsequent interaction of the energetic particles (electrons, $\gamma$-ray photons) propagating through the target. The details of this coupling scheme are summarized in the Methods section. 
 
 We consider a target of solid gold (size $ 2$ mm $\times 1$ cm $\times 1$ cm) with electron density $n_{0}\approx4.626\times 10^{24}$ cm$^{-3}$.
  It is illuminated by a linearly polarized laser pulse, striking the target at normal incidence, with wavelength $0.8\;\mu$m, intensity $10^{23}$ W/cm$^{2}$ and a Gaussian profile both in space and time. The spatial and temporal profiles have full-width-half-maximum $2\; \mu$m and $\tau=25$ fs respectively. These laser parameters are consistent with those expected at ELI-Nuclear Physics \cite{negoita2016laser}. A pre-plasma is assumed to exist in front of the target, consistent with a laser pre-pulse (as commonly preceeds the main high-intensity laser pulse) of intensity $\sim10^{14}$ W/cm$^{-2}$ and duration $\sim0.1$ ns. Although pre-pulses are usually considered a technological limitation in experiments, in our case the pre-pulse is advantageous: it substantially increases the laser absorption  by the plasma ($\phi_{tot}\sim25\%$) and thus the flux of energetic electrons and $\gamma$-rays going into the target.

A schematic representation of the main processes involved in ultra-intense laser illumination of heavy targets is shown in Fig.~\ref{disegno}.
 In the PIC simulation of the main pulse illuminating the target, the laser  propagates through the preplasma, reaching the relativistically-corrected critical density \cite{palaniyappan2012dynamics,kaw1970relativistic}, $3.7\times10^{23}$ cm$^{-3}$. Here the laser energy is partially absorbed and partially reflected. The reflected light `collides' with the laser-accelerated electrons that, consequently, radiate $\gamma$-rays photons by nonlinear Compton scattering. Contrary to Ref.~\cite{del2018efficient}, this process is particularly efficient ($\sim3.5\%$ in this case and  $\sim30\%$ for $I>10^{23}$ W/cm$^{2}$) because of the smooth density gradient in the pre-plasma.
 It was recently shown that laser-plasma interactions at a density close to the relativsitcally correceted critical density are optimal for gamma-ray production \cite{brady2012laser,brady2013gamma}.
 The propagation of the emitted $\gamma$-ray photons, together with laser-accelerated electrons, through the bulk of the target is then simulated using a Monte-Carlo code.

\begin{figure*}
\subfloat[\label{polar}]{%
        \centering
        \includegraphics[width=0.75\columnwidth]{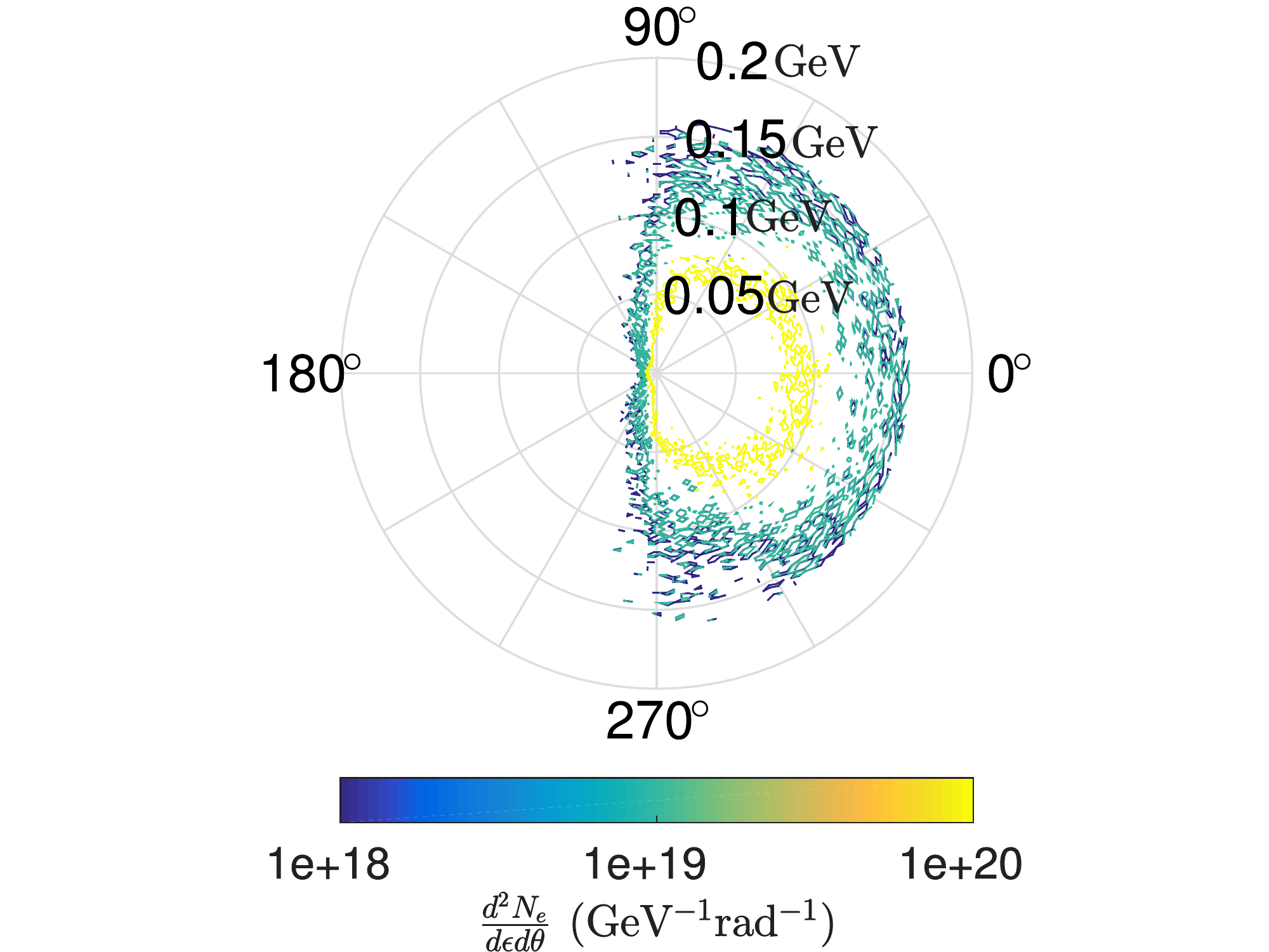}}
~
\subfloat[\label{polar_ph}]{%
        \centering
        \includegraphics[width=0.75\columnwidth]{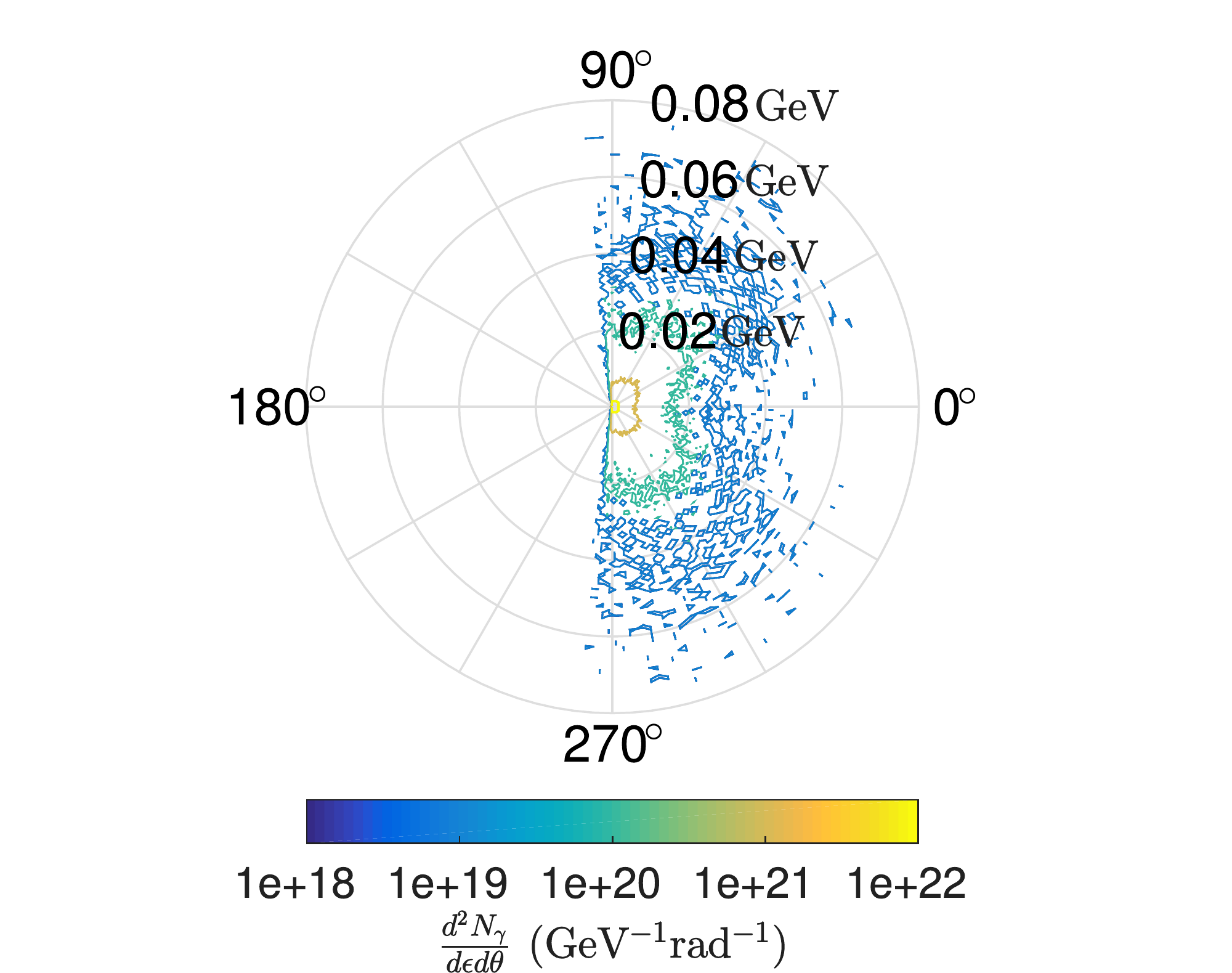}
                        }

      \caption{(a) Electron and (b) $\gamma$-ray distribution functions expressed as functions of energy (in GeV)  and direction of the velocity of the particles expressed as an angle between the velocity vector and laser propagation direction (in degrees) at $t=80$ fs. These distributions are integrated over space.
 }
\end{figure*}

In Fig.~\ref{disegno} we show the density of $\gamma$-ray photons at $t= 80$ fs, i.e.~when the interaction of the laser with the target has ended and most of the $\gamma$-ray photons have entered the bulk of the gold  ($t=0$ is when the main pulse strikes the target at $y=-13\;\mu$m). We see that $\gamma$-ray photons reflect the structure of the laser pulse that generated them: they are semi-circularly distributed in an annulus $\sim \tau c\approx 7.5\;\mu$m thick, where $c$ is the speed of light. 
The distribution functions of electrons and $\gamma$-rays at $t=80$ fs are shown in Figs.~\ref{polar} and \ref{polar_ph} respectively. From these figures we see that both the beams are  highly divergent, although it has recently been shown that the $\gamma$-ray divergence can be substantially reduced by lowering the target mass  \cite{capdessus2018relativistic}.
These electrons and photons then propagate into the bulk of the target. 

 The target is sufficiently large that away from the laser interaction region it remains essentially undisturbed by the laser pulse and therefore the propagation of the electrons and $\gamma$-rays shown in Figs. \ref{polar} and \ref{polar_ph} may be modelled with a Monte Carlo code \cite{agostinelli2003geant4,allison2006geant4}. Inside the target, $\gamma$-ray photons interact with gold nuclei, decaying into electron-positron pairs by the Bethe-Heitler process. In the same way, energetic electrons interact with the nuclei to radiate by bremsstrahlung, emitting $\gamma$-ray photons which can also decay to pairs by the Bethe-Heitler process. Each subsequent generation of these particles can also undergo these processes.

\begin{figure}
\subfloat[\label{23}]{%
        \centering
        \includegraphics[width=0.75\columnwidth]{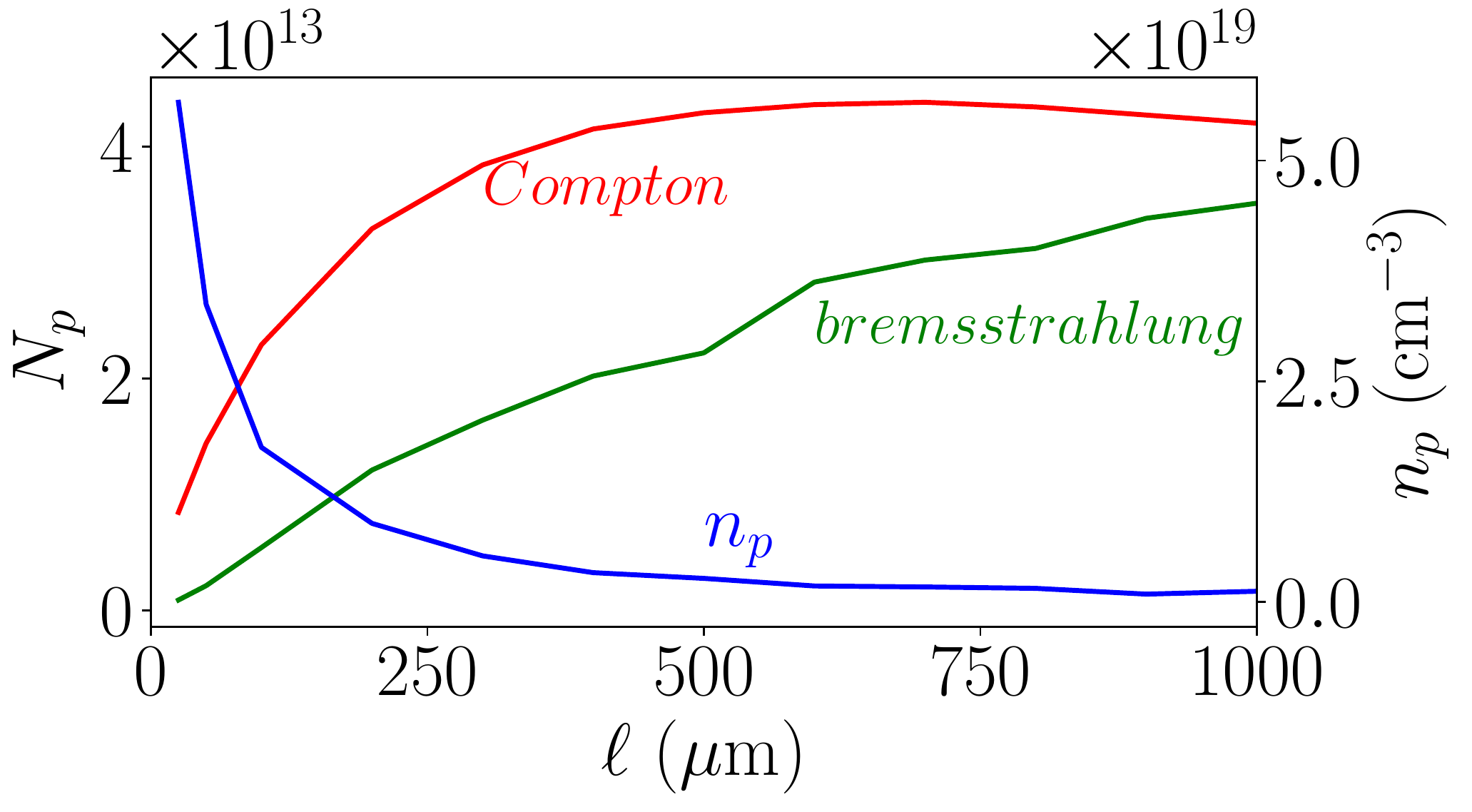}}

\subfloat[\label{I}]{%
        \centering
        \includegraphics[width=0.75\columnwidth]{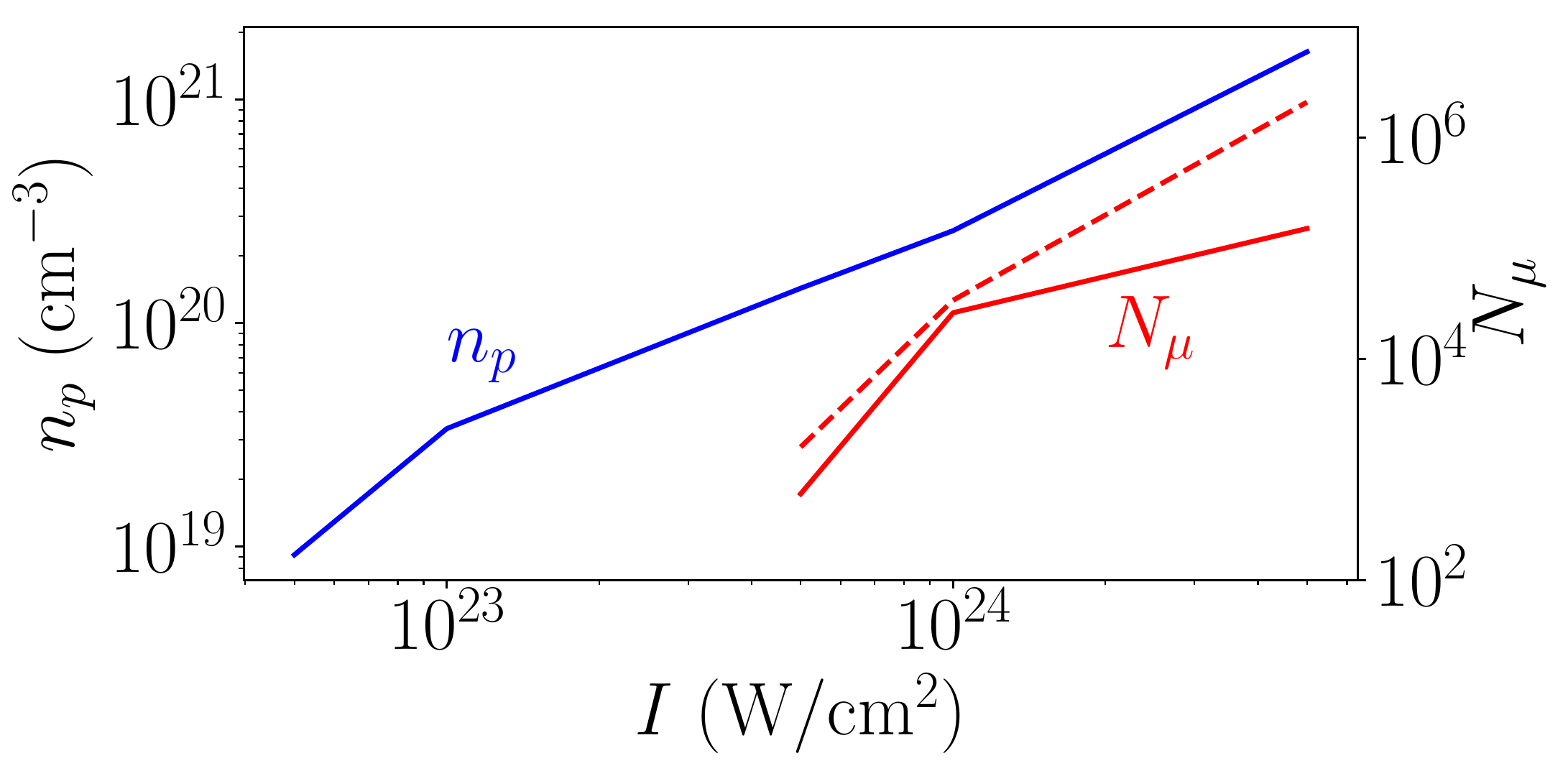}
                        }
      \caption{(a) Total number of positrons produced by bremsstrahlung (in green) and by nonlinear Compton scattering (in red) and the total peak density (in blue), computed at the rear side of the gold target, as a function of the target depth, for $I=10^{23}$ W/cm$^{2}$. (b) Positron peak density (blue) and total number of muon pairs produced (red) at the rear side of a 50 $\mu$m (continuous line) and a 500 $\mu$m (dashed line)  depth gold target as a function of laser intensity. }
\end{figure} 
 
 \textbf{Scaling with target thickness and laser intensity.}
Multiple Monte Carlo simulations have been performed in order to deduce the number of positrons produced and their density at the rear side of the gold target as a function of the target thickness $\ell$ (i.e. the length of the target in the direction of laser propagation). The results are shown in Fig.~\ref{23}. We see that pair production from $\gamma$-ray photons produced via non-linear Compton scattering occurs over a much shorter mean free path compared to pair production by $\gamma$-ray photons produced by bremsstrahlung. Thus thinner targets can be used for the production of the same number of photons, with consequences for the maximum achievable pair-plasma density.  Pair production saturates when the target is sufficiently thick that the later generations of emitted photons are of insufficinet energy to produce pairs.  For the channel via non-linear Compton scattering this occurs when $\ell\sim0.5$ mm. \color{black}

Figure \ref{23} shows the peak positron density in the electron-positron jets as they emerge from the rear  of the gold target, as a function of the target length. These density peaks are achieved  in cylindrical volumes of longitudinal size equal to the laser pulse duration multipled by the speed of light and radius 10 $\mu$m, shown in Fig.~\ref{disegno}. The jets are 3 to 4 orders of magnitude denser than previously achieved; so high to approach those of atmospheric gases ($\sim 10^{19}$ cm$^{-3}$) and exceed the non-relativistic critical density for commonly employed lasers, potentially enabling the study of dispersion of electromagnetic waves in pair plasmas. 
Because of the large divergence of the electron-positron jet, arising from the large divergence of the $\gamma$-rays emitted by nonlinear Compton scattering in the laser focus, the density tends to decrease as $\ell$ increases, while the total number of positrons continues to increase. As a result the bremsstrahlung channel is inefficient for generating dense electron-positron jets as it requires thicker targets. This highlights the importance of the new channel via non-linear Compton scattering.


\begin{table}
\begin{center}
\begin{tabular}{ c|ccccc} 

$I$ (W/cm$^{2}$) & $5\times10^{22} $ & $10^{23} $ & $5\times10^{23} $ & $10^{24} $ & $5\times10^{24} $  \\
\hline
$\phi_{tot}$ (\%) &  26 & 29  & 33 &35 & 42\\ 

$\phi_{e}$ (\%)  &25 & 25 & 16 &13 & 6.6\\ 
$\phi_{\gamma}$ (\%) & 1.3 & 3.5 & 17 & 22 & 32\\ 
$\phi_{p}$ (\%) & 0 & 0 & 0 & 0.1 & 2.9\\ 
$<\epsilon_{p}>$ (MeV) &4.3 & 6 &15&37&250\\ 
$n_{p}$ ($10^{19}$ cm$^{-3}$) & $0.92$ & $3.4$ & $14$& $26$& $163$\\ 
$\delta_{sd}$ ($\mu$m) & 3.8 & 2.3 & 1.7&2 & 2\\

\end{tabular}
\end{center}
\caption{Scaling of main quantities with laser intensity: the efficiency ($\phi$) of laser conversion into particles before nuclear interactions ($e$, $\gamma$ and $p$ subscripts refer to electrons, $\gamma$-ray photons and positrons), average positron energy $<\epsilon_{p}>$, peak density ($n_{p}$) and electron-positron plasma relativistic skin depth ($\delta_{sd} = \sqrt{<\epsilon_{p}>/8\pi n_{p} e^{2}}$, with $e$ as the elementary charge) for $\ell=50\mu m$. 
\label{tabella}}
\end{table}

The investigation performed above has been repeated for different laser intensities, from $I=5\times10^{22}$ to $5\times10^{24}$ W/cm$^{2}$, leaving all the other parameters unmodified. 
Here the target thickness was fixed $\ell=50 \;\mu$m.

Table \ref{tabella} summarizes how the main quantities are modified by laser intensity. 
As the intensity increases, so does the efficiency of the laser conversion to $\gamma$-ray photons by nonlinear Compton scattering, increasing the number of positrons produced in the target. For $I\gtrsim10^{24}$,  pair creation by the alternative Breit-Wheeler process  \cite{burke1997positron} in the laser interaction region also plays a role \cite{del2018efficient,slade2018identifying,grismayer2015seeded,kirk2009pair}, although such high intensities are unlikely to be reached in the near term.  As a result of increased positron production, the relativistic collisionless skin depth in the pair-jet emitted at the rear of the target decreases with increasing laser intensity. Since the density peak is computed in a cylinder of radius $\sim10\;\mu$m, we deduce that the scale of the emitted jets is $\sim5$ times larger than the relativistic skin depth $\delta_{sd}$, while the Debye sphere ($\sim4/3\pi\delta_{sd}^{3}$) is composed by $\sim10^{8}$ particles. 

 \textbf{Producing neutral electron-positron and muon antimuon jets.}
This significant and efficient new channel enables the laboratory study of collective effects in relativistic pair plasma jets.
For example, these pair plasma jets are appealing for the laboratory investigation of Weibel current filamentation, which is considered responsible of the generation of ultra-bright $\gamma$-ray emission in astrophysical environments \cite{mirabel1999sources}. Indeed, the complete manifestation of this instability requires that the plasma scale length is larger than the relativistic skin depth \cite{spitkovsky2008particle}, a condition which has so far proved to be experimentally elusive.

The scaling of the peak positron density at the rear of the gold target as a function of laser intensity (of the main pulse) is shown in Fig.~\ref{I}. A similar number of pair-produced electrons is expected at the rear of the target.  However, electrons: (i) accelerated in the laser-produced plasma and (ii) resulting from ionization in the bulk target also reach the target rear.  These electrons break the charge neutrality of the electron-positron plasma at the target rear, essential for the investigation of collective processes in a pair-plasma. These excess electrons could be removed by using a more sophisticated target configuration, for example by employing a vacuum gap between two targets where the sheath field at the rear of the first would confine electrons \cite{macchi2013ion,link2011effects}.

Photons generated by non-linear Compton scattering with energies exceeding 100 MeV can result in muon-antimuon pair production. 
 Figure \ref{I} shows that above $5\times10^{23}$ W/cm$^{2}$ copious numbers of muons -- $10^3$-$10^6$ -- are emitted at the rear of the gold target.
The possibility of prolifically producing muons with laser light provides an alternative to current sources of these particles, with intriguing applications, such as muon cathalized fusion \cite{breunlich1989muon} and muon colliders for neutrino factories, investigated within the Muon Accelerator Program \cite{palmer2013overview}. A high flux muon source may also be used for calibrating detectors, and for muon radiography -- a technique also applied in archaeology \cite{morishima2017discovery}.


In conclusion, we presented a new mechanism for the production of dense pair-jets, using multi-PW lasers. Moving to higher intensity lasers introduces new physical processes, such as nonlinear Compton scattering in the laser-produced plasma. This is a very efficient generator of high energy photons
which can decay to a dense pair-jet by the Bethe-Heitler process in a dense solid target. Electron-positron plasmas with densities above the atmospheric density and in volumes where collective effects are important are predicted in the laboratory for the first time.  This density also exceeds that required to produce a Bose-Einstein condensate, although this requires liquid helium temperatures \cite{liang1988laser,cassidy2007physics}. Our findings show this process is not limited to electron-positron pairs production as the nonlinear Compton scattering 
followed by the analogous Bethe-Heitler process for muon production will also produce copious muon-antimuon pairs.

 \section*{Methods}
 
\textbf{Numerical modeling.}
To investigate the interaction of high intensity lasers with thick heavy targets we have combined three different types of modeling: (i) prepulse ablation is analytically modeled; (ii) the interaction of the main laser-pulse with the target is modeled using the QED particle-in-cell code EPOCH, initialized with the ablated density from (i); (iii) the energetic particle propagation through the bulk of the heavy target is described with the Monte Carlo code GEANT4, initialized with the particle (electrons, positrons, $\gamma$-rays) distribution functions obtained from (ii).

The pre-pulse ablates the front of the target, producing a low-density preplasma.
Its  effect has been modeled at time $t$ by assuming that the ablation takes place in a time interval equal to the laser prepulse duration $\tau_{p}$ and that the ablated material moves at the sound speed $c_{s}$. Therefore, when the main pulse hits the target, the target density is:
\begin{equation}
n_{e}=
\begin{cases}
n_{0} \exp\left({\frac{y}{c_{s}\tau_{p}}}\right)\; & \mbox{if} \; \mbox{y}<0\\
n_{0}\; & \mbox{if} \; \mbox{y}>0
\end{cases}.\label{ablation density}
\end{equation} 
The sound speed, obtained by assuming an ideal gas equation of state for the target and a laser prepulse intensity $I\sim10^{14}$ W/cm$^{-2}$ (giving a temperature of $T\approx25$ eV), was $c_{s}\approx38$ km/s.   Slightly varying $c_{s}$ does not affect the laser conversion efficiency strongly.

The laser interaction region was simulated using EPOCH \cite{arber2015contemporary}. 
EPOCH is a particle-in-cell code: the plasma is described as a collection of macroparticles (electrons, ions and, eventually, positrons and $\gamma$-ray photons), with a statistical weight, that move according to the Lorentz force law under the action of the electromagnetic fields in the plasma. These electromagnetic fields, due to the laser and also collective plasma motion, are discretized on a spatial grid and evolve according to Maxwell's equations.

EPOCH also accounts for non-linear Compton scattering and multiphoton Breit-Wheeler pair production \cite{ridgers2014modelling}. These are modeled using a now standard quasi-classical description \cite{baier1968processes}. The electrons, positrons and $\gamma$-ray photons move classically between pointlike emission events.   This is valid when the photon (and electron-positron pair) formation length is small, i.e. the emission occurs over a very small region, as is the case for intensities $\gg10^{18}$\ Wcm$^{-2}$ considered here \cite{kirk2009pair}.  The emission rates are approximated by the well-known rates in constant crossed electric and magnetic fields provided the electromagnetic fields in the laboratory frame are well below the critical (Schwinger) field, as is also the case here.

The propagation of energetic particles inside the unperturbed bulk of the gold target has been modeled using the code GEANT4  \cite{agostinelli2003geant4,allison2006geant4}. 
This is a Monte-Carlo simulation code that is used to model radiation production and transport. A number of physics libraries are available and for the present work `Lawrence Livermore'  \cite{apostolakis2015progress} (https://geant4.web.cern.ch/node/1619) was employed due to the energy regime. It includes bremsstrahlung and Bethe-Heitler pair production (electron-positron and dimuon).

The coupling of EPOCH, for simulating the laser interaction region, to GEANT4 for simulating the propagation of the particles and radiation through the bulk of the target, was done as follows.  The distribution function of electrons, positrons and $\gamma$-ray photons  was recorded and integrated spatially after 80 fs 
from when the main pulse illuminated the target at $y=13\;\mu$m.
 GEANT4 was initialised with these distributions.  A $180^{\circ}$ angular divergence and an initial radius of 5 $\mu$m was applied to the particles injected into the GEANT4 simulations, consistent with the distributions observed in the EPOCH simulations.
To link the two dimensional EPOCH simulation to the three dimensional GEANT4 simulation, all the results derived from the EPOCH simulations have been interpreted as three dimensional simulations in cylindrical geometry. E.g.~the total number of particles in the EPOCH simulation was expressed in units of meters along the z-direction and was rescaled by multiplying by the factor $\pi/4d_{FWHM}$, where $d_{FWHM}$ is the full-width-half-maximum of the laser pulse profile.

\textbf{Numerical parameters.} All the EPOCH simulations performed have been run in two dimensions (planar geometry). They have been initialized using $600\times600$ spatial grid cells to describe the simulation box $[-20\;\mu\mbox{m},20\;\mu\mbox{m}]\times [-13\;\mu\mbox{m},20\;\mu\mbox{m}]$. The portion of the target contained in this region was represented by 48 million macroelectrons and the same number of macroions. Periodic boundary conditions have been used in the x-direction while transmission has been allowed along the y-direction (the laser propagated in the $y$-direction). The interval y $\leq -12\;\mu$m was initialized void of macroparticles.

All GEANT4 simulations were run for $10^{8}$ events, a suitable compromise between computing time and particle production for the physics processes with a small cross section. Photons from EPOCH with energies below twice the electron rest-mass were not considered, the results were normalized and scaled for consistency with the EPOCH simulations to determine the number of electron-positron pairs. In the case of muon antimuon pairs the production cross section was weighted in order to produce a statistically valid result. Again, the results are normalised and scaled to the EPOCH calculations to determine the total number of muon antimuon pairs produced.


\section*{Acknowledgments}
This work was funded by the UK Engineering and Physical Sciences Research Council (EP/M018156/1).


 \end{document}